\documentclass[twocolumn,showpacs,amsmath,amssymb,superscriptadress,prc]{revtex4-1}
\usepackage{epsfig,amsmath,graphicx,amssymb,tabularx}
\usepackage[dvipdfmx,bookmarks=true,colorlinks,citecolor=blue,linkcolor=blue,anchorcolor=blue,filecolor=blue,urlcolor=blue]{hyperref}
\usepackage{CJK}
\usepackage{dcolumn}% Align table columns on decimal point
\usepackage{bm}% bold math

\def\beq{\begin{equation}}
\def\eeq{\end{equation}}
\def\bea{\begin{eqnarray}}
\def\eea{\end{eqnarray}}
\begin{document}
%\begin{CJK*}{GBK}{}

\title{Dissipation dynamics and spin-orbit force in time-dependent Hartree-Fock theory}

%\author{Gao-Feng Dai~(´÷¸ß·å)$^1$}
%\author{Lu Guo~(¹ùè´)$^{2,1}$}
%\email{luguo@ucas.ac.cn}
%\author{En-Guang Zhao~(ÕÔ¶÷¹ã)$^{1,3}$}
%\author{Shan-Gui Zhou~(ÖÜÉƹó)$^{1,3}$}
%\email{sgzhou@itp.ac.cn}

\author{Gao-Feng Dai$^1$}
\author{Lu Guo$^{2,1}$}
\email{luguo@ucas.ac.cn}
\author{En-Guang Zhao$^{1,3}$}
\author{Shan-Gui Zhou$^{1,3}$}

\affiliation{$^1$ State Key Laboratory of Theoretical Physics, Institute of Theoretical Physics, Chinese Academy of Sciences, Beijing 100190, China}
\affiliation{$^2$ School of Physics, University of Chinese Academy of Sciences, Beijing 100049, China}
\affiliation{$^3$ Center of Theoretical Nuclear Physics, National Laboratory of Heavy Ion Accelerator, Lanzhou 730000, China}

\date{\today}

\begin{abstract}
We investigate the one-body dissipation dynamics in heavy-ion collisions of $^{16}{\rm O}$+$^{16}{\rm O}$ using a fully three-dimensional time-dependent
Hartree-Fock (TDHF) theory
with the modern Skyrme energy functional and without any symmetry restrictions. The energy dissipation is revealed to decrease in deep-inelastic
collisions of the light systems as the bombarding energy increases owing to the competition between collective motion and single-particle degrees of freedom.
The role of spin-orbit force is given particular emphasis in deep-inelastic collisions. The spin-orbit force causes a significant enhancement of the dissipation. The
time-even coupling of spin-orbit force plays a dominant role at low energies, while the influence of time-odd terms is notable at high energies.
About 40-65\% of the total dissipation depending on the different parameter sets is predicted to arise from the spin-orbit force. The theoretical fusion
cross section has a reasonably good agreement with the experimental data, considering that no free parameters are adjusted to reaction dynamics in the TDHF approach.
\end{abstract}


\maketitle
%\end{CJK*}

%%%%%%%%%%%%%%%%%%%%%%%%%%%%%%%%%%%%%%%%%%%%%%%%%%%%%

\section{Introduction}
\label{modification:Intro delete}
Time-dependent Hartree-Fock (TDHF) theory, originally proposed by Dirac~\cite{Dirac1930_PCPS26-376}, has been widely applied to describe the fusion excitation functions, deep-inelastic collisions, fission, collective excitation, and nuclear molecular resonances; for reviews see Refs.~\cite{Negele1982_RMP54-913,
Simenel2012_EPJA48-152}. It provides, from a fully microscopic point of view,
the dynamical foundation of large amplitude collective motion~\cite{Boneche1976_PRC13-1226}.
TDHF theory has exactly treated the one-body dissipation caused by the collisions of nucleons with mean-field potential and nucleon transfer between the colliding partners. The two-body dissipation, which was included in quantum molecular dynamics models~\cite{Aichelin1991_PR202-233, Wang2002_PRC65-064608, Wen2013_PRL111-012501,*Wen2014_arXiv1407.5912} and is likely
to play a significant role at high energies, has been neglected within the framework of mean-field dynamics.

\label{modification:Intro Para2}
A number of approaches based on TDHF have been developed to investigate the dissipation mechanism in heavy-ion collisions and resonance dynamics.
Density-constrained TDHF (DC-TDHF)~\cite{Umar2006_PRC74-021601} has utilized the time-dependent instantaneous densities as the constraints to perform the static Hartree-Fock minimization, thus capable of extracting the excitation energy~\cite{Umar2009_PRC80-041601} and the underlying nucleus-nucleus interaction potential~\cite{Umar2006_PRC74-021601}. The method is the dynamical analog of static adiabatic approximation and has been applied to calculate the fusion cross section at both sub- and above-barrier energies~\cite{Umar2010_PRC81-064607,Umar2012_PRC85-055801,Simenel2013_PRC88-024617}.
In the dissipative dynamics TDHF (DD-TDHF) approach~\cite{Washiyama2008_PRC78-024610,Washiyama2009_PRC79-024609}, mean-field evolution was assumed to be properly reduced to one-dimensional dissipative dynamics. By using the macroscopic reduction procedure, the one-body
dissipation in the entrance channel of heavy-ion fusion reactions has been extracted from the simulations of microscopic TDHF~\cite{Washiyama2008_PRC78-024610,
Washiyama2009_PRC79-024609}. The technique of Wigner distribution function~\cite{Loebl2011_PRC84-034608,loebl2012_PRC86-024608} has
been applied to probe the single-particle dissipation in TDHF from a phase-space perspective. As indicated by Loebl and collaborators, the one-body dissipation is not enough to achieve the true equilibrium as the reaction proceeds to longer contact times~\cite{Loebl2011_PRC84-034608}. The tensor force has been included in recent TDHF calculations~\cite{Iwata2011_PRC84-014616,Dai2014_SciChinaPMA57-1618,Fracasso2012_PRC86-044303}.
The time-even contribution
of the tensor force was found to have important roles in heavy-ion collisions~\cite{Iwata2011_PRC84-014616,Dai2014_SciChinaPMA57-1618}. The full tensor expression was introduced to study the resonance dynamics~\cite{Fracasso2012_PRC86-044303}.

\label{modification:Intro Para3}
Since TDHF theory is based on the independent-particle approximation, the expectation value of one-body observables can be well described.
However, the description of dynamical fluctuation and two-body dissipation requires the theoretical development beyond TDHF. The stochastic extension of TDHF theory~\cite{Ayik2009_PRC79-054606,Washiyama2009_PRC80-031602} has been proposed to include the dynamical fluctuations of collective variables beyond mean field in a fully microscopic framework. The Balian-V\'en\'eroni
variational principle~\cite{Balian1984_PLB136-301,Broomfield2008_JPG35-095102} was first applied to realistic calculations of fragment mass and charge distributions in heavy-ion collisions by Simenel~\cite{Simenel2011_PRL106-112502}. By comparing with the standard TDHF calculations,
an increase of the fluctuations has been found and the agreement with experimental data has been much improved. The two-body dissipation arising from nucleon-nucleon collisions has been taken into account in a quantum mechanical way with
the extended TDHF~\cite{Lacroix2004_PPNP52-497,Assi2009_PRL102-202501} or time-dependent density matrix (TDDM) approaches ~\cite{Tohyama1987_PRC36-187,Tohyama2001_PLB516-415,Tohyama2002_PRC65-037601}. So far only a few studies on the dynamical fluctuation and two-body dissipation have been implemented because beyond TDHF calculations require a lot of numerical efforts and computational time.

\label{modification:Intro Landau-Zener}
The dissipation dynamics has an influence on the behavior of nucleus-nucleus collision.
The energy dissipation, closely associated with the interplay between the collective motion and the single-particle motion, depends on the delicate
balance between reaction time and rearrangement time of the mean field. When the collective motion is slow enough so that the mean field has enough time to
rearrange itself, the reaction dynamics follows the adiabatic motion.
For example, the experimentally observed Landau-Zener effect~\cite{Landau1932_PZS2-46, Zener1932_PRSLSA137-696} in inelastic cross section is due to the breakdown of adiabatic condition. The Landau-Zener effect and its applications to heavy-ion collisions were, for the first time, investigated with TDHF theory by constructing the boost-invariant adiabatic single-particle states~\cite{Guo2007_PRC76-014601}.

\label{modification:Intro spin-oribt}
Earlier TDHF calculations systematically underestimated the dissipation of energy from the relative collective motion into internal degrees of freedom
due to the simplified effective interaction and symmetry restrictions~\cite{Krieger1978_PRC18-2567, Davies1978_PRC18-2631, Davies1981_PRC23-2042, Maruhn1985_PRC31-1289}. For example, there has been the puzzle of the fusion window anomaly, which was later solved by including the time-even terms of spin-orbit force in TDHF calculations~\cite{Umar1986_PRL56-2793}.
The fusion excitation function was enhanced by as much as 20\% for the $^{40}$Ca+$^{40}$Ca system by removing the isospin symmetry restriction on the nuclear wave functions~\cite{Krieger1978_PRC18-2567}.
In Ref.~\cite{Reinhard1988_PRC37-1026} it was demonstrated that, despite using various approximations,
the time-even spin-orbit force has a significant effect on the dissipation dynamics in heavy-ion scattering.
However, the time-even spin-orbit interaction and the three-dimensional geometry have not been incorporated simultaneously
in these earlier TDHF calculations, and also the time-odd terms of spin-orbit force have been neglected.
The first three-dimensional TDHF calculations including the full spin-orbit interaction~\cite{Kim1997_JPG23-1267,Simenel2001_PRL86-2971} have been
performed using the TDHF3D code. The role of spin-orbit interaction on single-particle states in heavy-ion collisions was discussed in Ref.~\cite{iwata2008_IJMPE17-1660}.
The time-odd terms of spin-orbit force was addressed to be the origin of the spin excitation mechanism~\cite{Maruhn2006_PRC74-027601}.

With the development of computer ability, a fully three-dimensional TDHF calculation with modern energy functional and
without any symmetry restrictions has been realized in recent years. It is expected to provide a better description of heavy-ion collisions~\cite{Umar2006_PRC73-054607, Guo2008_PRC77-041301, Simenel2010_PRL105-192701,
Guo2012_EPJWoC38-09003,Simenel2013_PRC88-064604,Sekizawa2013_PRC88-014614,Simenel2014_PRC89-031601,Umar2014_PRC89-034611}
and resonance dynamics~\cite{Simenel2003_PRC68-024302, Nakatsukasa2005_PRC71-024301, Maruhn2005_PRC71-064328, Umar2005_PRC71-034314, Reinhard2007_EPJA32-19,Simenel2009_PRC80-064309}.
The purpose of present article is to systematically investigate the dissipation dynamics in deep-inelastic collisions and the role of spin-orbit force
by using modern TDHF calculations.
The article is organized as follows. In Sec.~\ref{theory}, we briefly explain TDHF theory and the details in numerical calculations. In Sec.~\ref{discuss},
we illustrate the dissipation dynamics and the role of time-even and time-odd terms
of spin-orbit force. The fusion cross section obtained from modern TDHF calculations is compared with experimental data. A summary is given in Sec.~\ref{summary}.

\section{the time-dependent Hartree-Fock approach}
\label{theory}
\label{modification:Theory Skyrme}
Most TDHF calculations employ Skyrme effective interaction~\cite{Skyrme1956_PM1-1043}. The Skyrme energy functional provides a good description of nuclear ground state properties and collective excitations.
Various parametrizations of Skyrme effective interaction have been fitted with a different emphasis on nuclear structure properties. There are no free parameters adjusted on the reaction dynamics. In order to see the parameter dependence, three Skyrme parametrizations of SLy4~\cite{Chabanat1998_NPA635-231,*Chabanat1998_NPA643-441}, SkM*~\cite{Bartel1982_NPA386-79}, and UNEDF1~\cite{Kortelainen2012_PRC85-024304} have been implemented to study the dissipation dynamics in heavy-ion collisions in this work.
We will not show the calculations with other parametrizations since they do not produce a significant difference for the purpose of the present work.

The time evolution of self-consistent mean-field generated by all the particles is described by the TDHF equation
\beq
i\hbar\partial_t\psi_\alpha=h(\rho, \tau, \vec\sigma, \vec j, \vec J)\psi_\alpha,
\eeq
where the time-dependent single-particle Hamiltonian $h$ consists of density $\rho$, kinetic density $\tau$, spin density $\vec\sigma$,
current $\vec j$, and spin-orbit density $\vec J$~\cite{Engel1975_NPA249-215}. The main approximation of TDHF theory
is to treat the many-body wave function as an independent particle state at any time. Starting from a proper initial state obtained from the static HF equation, the TDHF equation is solved to determine the wave functions at each time of
dynamical evolution. The mean-field Hamiltonian $h$ is derived from the energy density functional (EDF)
\beq
E=\int d^3r \mathcal H(\rho, \tau, \vec\sigma, \vec j, \vec J),
\eeq
by using the time-dependent variation
\label{modification:Theory equation}
\beq
h_{ij}=\frac{\partial E}{\partial \rho_{ji}}.
\eeq
The energy functional $\mathcal H$ consists of free kinetic energy, Skyrme interaction, and Coulomb energy with exchange in the Slater approximation.
Note that the TDHF approach treats the static properties and reaction dynamics in a unified theoretical framework and the same energy functional.

For the sake of later discussions, the time-even and time-odd terms of spin-orbit force
\beq
\mathcal H_{\rm ls}^{\rm even} = -\frac{1}{2}t_4\left(\rho\nabla\cdot\vec J+\sum_q\rho_q\nabla\cdot\vec J_q\right),
\label{Heven}
\eeq
\beq
\mathcal H_{\rm ls}^{\rm odd} = -\frac{1}{2}t_4 \left(\vec s\cdot\nabla\times\vec j + \sum_q \vec s_q\cdot\nabla\times\vec j_q\right),
\label{Hodd}
\eeq
are expressed in terms of various densities and Skyrme parameter $t_4$. The index $q$ denotes protons and neutrons.
The inclusion of both time-even and time-odd terms guarantees the Galilean invariance which should be met in a meaningful theory of heavy-ion collisions~\cite{Engel1975_NPA249-215}. The odd-odd terms of spin-orbit force are important to assure the energy conservation in the free translation of a nucleus over the grids as reported in Ref.~\cite{Maruhn2006_PRC74-027601}.

The set of nonlinear TDHF equations is solved on a three-dimensional Cartesian coordinate-space without any symmetry restrictions. The derivative
is solved with fast Fourier transformation in Fourier representation. The numerical box of coordinate space is 32$\times$24$\times$24 $\rm fm^{3}$ for the present calculations
of the light system $^{16}$O+$^{16}$O. The colliding nuclei locate at an initial center of mass distance of 16 fm. Before reaching
this distance, the colliding nuclei follow the Rutherford trajectory. The grid spacing is taken as 1 fm and a time step of dynamical evolution as 0.2 fm/c.
The dynamical unitary propagator is expanded up to the sixth order of Taylor expansion. The choice of these
parameters guarantees good numerical accuracy during the dynamical evolution for all the cases studied here.
The shift of particle number and total energy is less than, respectively, 0.01 and 0.1 MeV during the time evolution.

\label{modification:Theory transparency}
Fusion occurs when the collective kinetic energy is entirely converted into the internal excitation of a well-defined compound nucleus.
TDHF fusion cross section is calculated at each energy by the sharp-cutoff approximation~\cite{Bonche1978_PRC17-1700}
\beq
\sigma_{\rm fus}=\frac{\pi\hbar^2}{2\mu E_{\rm c.m.}}[(l_{\rm max}+1)^2-(l_{\rm min}+1)^2],
\eeq
where $\mu$ is the reduced mass of the system and $E_{\rm c.m.}$ is the initial center-of-mass energy. The quantities $l_{\rm max}$ and $l_{\rm min}$ denote
the maximum and minimum orbital angular momentum for which fusion happens. At low collision energy, the minimum orbital angular momentum is zero.
As the increase of incident energy, there usually appears the nonzero lower limit of orbital angular momentum due to the transparency behavior in central collisions~\cite{Davies1980_PRL44-23,Devi1981_PRC23-1064,Maruhn1985_PRC31-1289}.

\section{results and discussions}
\label{discuss}
\label{modification:result groups}
We have performed fully three-dimensional TDHF calculations using Sky3D code~\cite{Maruhn2014_CPC185-2195}
for heavy-ion collisions of $^{16}$O+$^{16}$O
with modern Skyrme EDF and without any symmetry restrictions.
This system has been studied with TDHF by several groups on fusion, deep-inelastic collision, the associated dissipation mechanisms, and the
extensive comparison with experimental data~\cite{Bonche1978_PRC17-1700,Maruhn1985_PRC31-1289,Umar1986_PRL56-2793,Reinhard1988_PRC37-1026, Tohyama2002_PRC65-037601,Maruhn2006_PRC74-027601,Umar2006_PRC73-054607,Umar2006_PRC74-021601,Guo2007_PRC76-014601,Guo2008_PRC77-041301,Washiyama2008_PRC78-024610,Umar2009_PRC80-041601,
Guo2012_EPJWoC38-09003,Loebl2011_PRC84-034608,Simenel2013_PRC88-024617,Dai2014_SciChinaPMA57-1618}.
The present studies will be compared with the earlier calculations and the available experimental data in order to clarify the impact of new terms in Skyrme EDF
and the geometric symmetries on the reaction dynamics.

\begin{figure}
  \includegraphics[width=280pt]{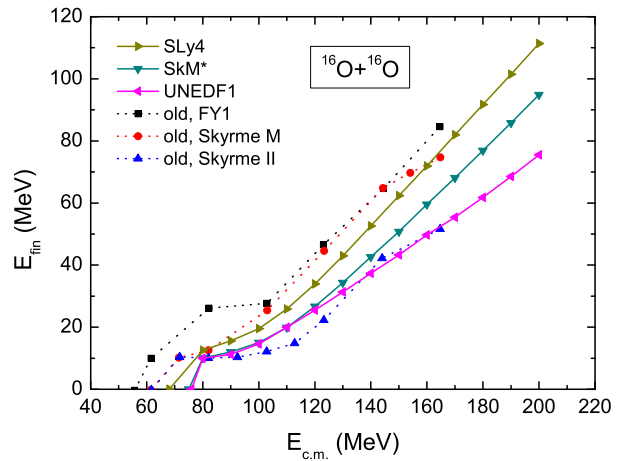}\\
  \caption{(Color online) Final relative kinetic energy as a function of center-of-mass energy for head-on collisions of $^{16}$O+$^{16}$O. The solid lines
  show the results of present studies using three Skyrme parametrizations, and the dashed lines are taken from Ref.~\cite{Reinhard1988_PRC37-1026} for comparison.}
  \label{Efin}
\end{figure}

One of the most interesting experimental observables in heavy-ion scattering is the relative kinetic energy of the separating ions.
Figure~\ref{Efin} shows the final relative kinetic energy as a function of initial center-of-mass (c.m.) energy for head-on collisions
of $^{16}$O+$^{16}$O. The solid lines with different symbols denote the present studies using three Skyrme parametrizations SLy4, SkM*, and UNEDF1. For comparison the dashed lines from earlier TDHF calculations~\cite{Reinhard1988_PRC37-1026} are also displayed.
The difference between present studies and earlier calculations mainly lies in three points. First, the axial symmetry was assumed in earlier calculations,
while the present study does not impose any symmetry restrictions. Second, we include the full spin-orbit (l*s) force,
and yet only time-even terms of l*s force were incorporated in earlier studies. Third, the terms involving the gradient of density
in Skyrme EDF were replaced with the finite-range Yukawa-folding form in earlier studies to simplify the numerical calculations, compared with the full treatment in present studies.

\label{modification:result two-body collisions}
The heavy-ion scattering behavior studied here is for energies from the threshold of inelastic scattering to an initial c.m. energy of 200 MeV.
It should be noted that TDHF dynamics can only be used to discuss the one-body dissipative mechanism.
In the energy range of the present study, the two-body collision terms are likely to play a significant role. However, as presented in the Introduction, beyond TDHF calculations still remain a numerical difficulty. We then present this work as a study of one-body dissipation dynamics.

As seen in Fig.~\ref{Efin}, the overall
trend of solid lines is similar for the three parametrizations, and yet there gradually appears a distinct discrepancy
as the increase of c.m. energy. In the energy range studied here, the outgoing ions carry the largest kinetic energies
using SLy4, and the smallest with the newly fitted parametrization UNEDF1. We may therefore expect that the energy dissipation will be the most distinguished with UNEDF1, and the weakest using SLy4 among the three parametrizations. The dashed lines from the earlier studies are shifted notably depending on the Skyrme parametrizations. By comparing the results of the
present and earlier studies,
the sensitivity of earlier calculations on Skyrme parametrizations is attributed to the approximations stated above on the effective interaction and
geometric symmetry. It should be noted that the final energy has an uncertainty of 2-3 MeV because the nucleons emitted from the highly excited fragments are reflected from the boundaries of the numerical box, as seen in deep-inelastic collisions~\cite{Guo2008_PRC77-041301} and giant resonance calculations~\cite{Reinhard2006_PRE73-036709}. This will not cause an essential distinction for the purpose of present studies.
In the following, the results with SLy4 will be taken as an example for further discussions.

\begin{figure}
  \includegraphics[width=280pt]{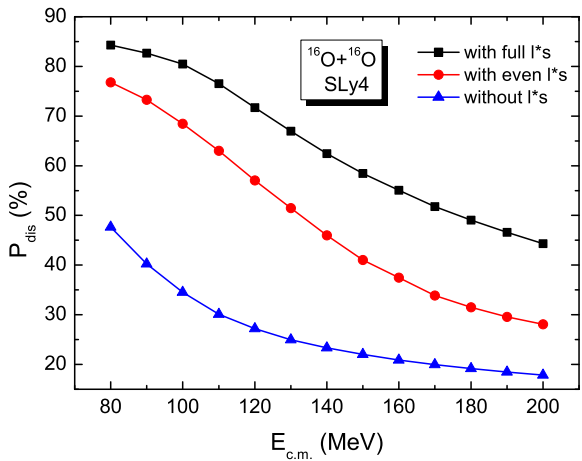}\\
  \caption{(Color online) Percentage of energy dissipation as a function of center-of-mass energy for head-on collisions of $^{16}$O+$^{16}$O with parametrization SLy4.
  The black, red, and blue lines represent the TDHF calculations involving full l*s, time-even l*s,
  and no l*s force. }
  \label{Pdis}
\end{figure}

\begin{figure}
\includegraphics[width=120pt]{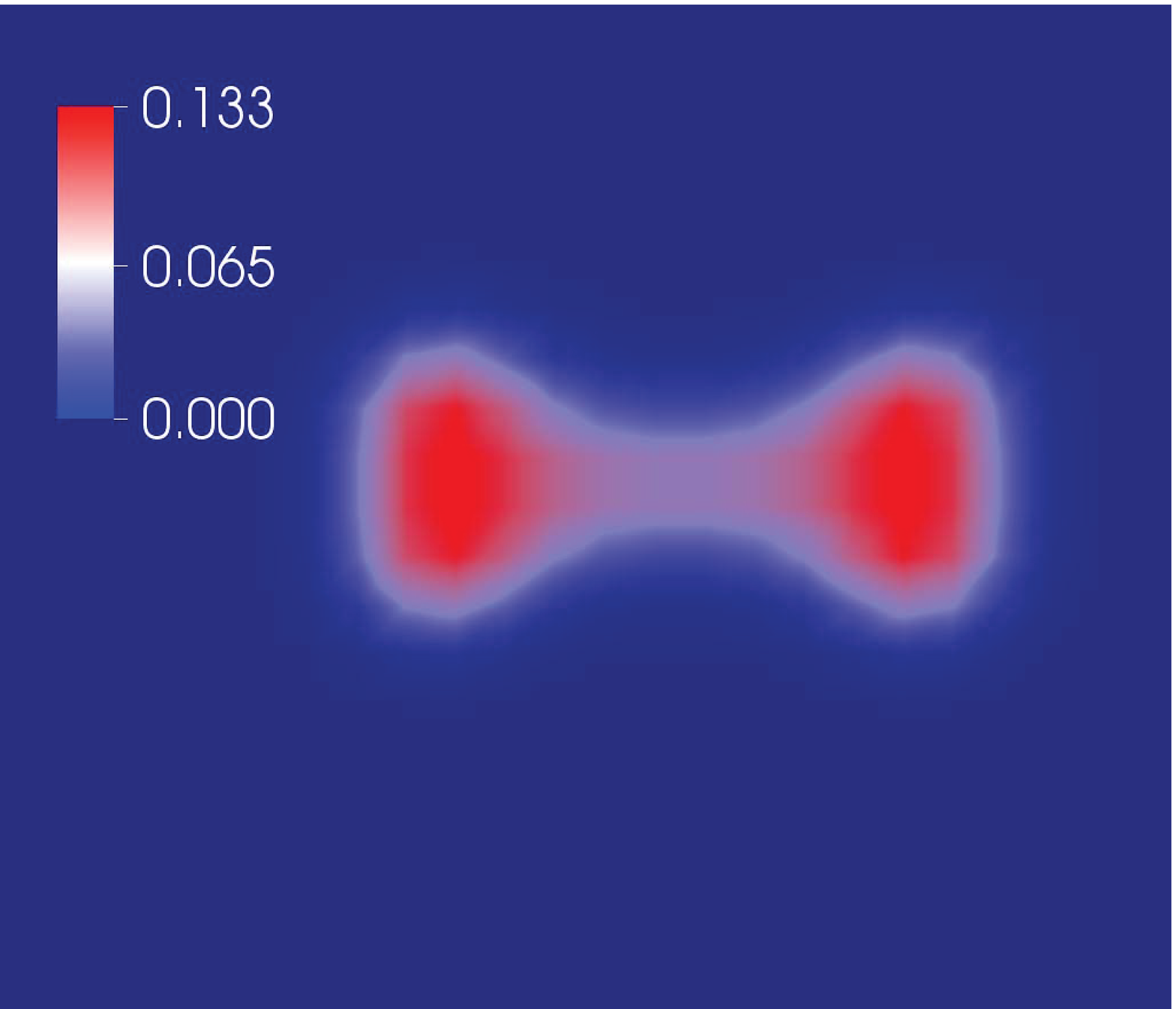}\\[-36pt]
\includegraphics[width=120pt]{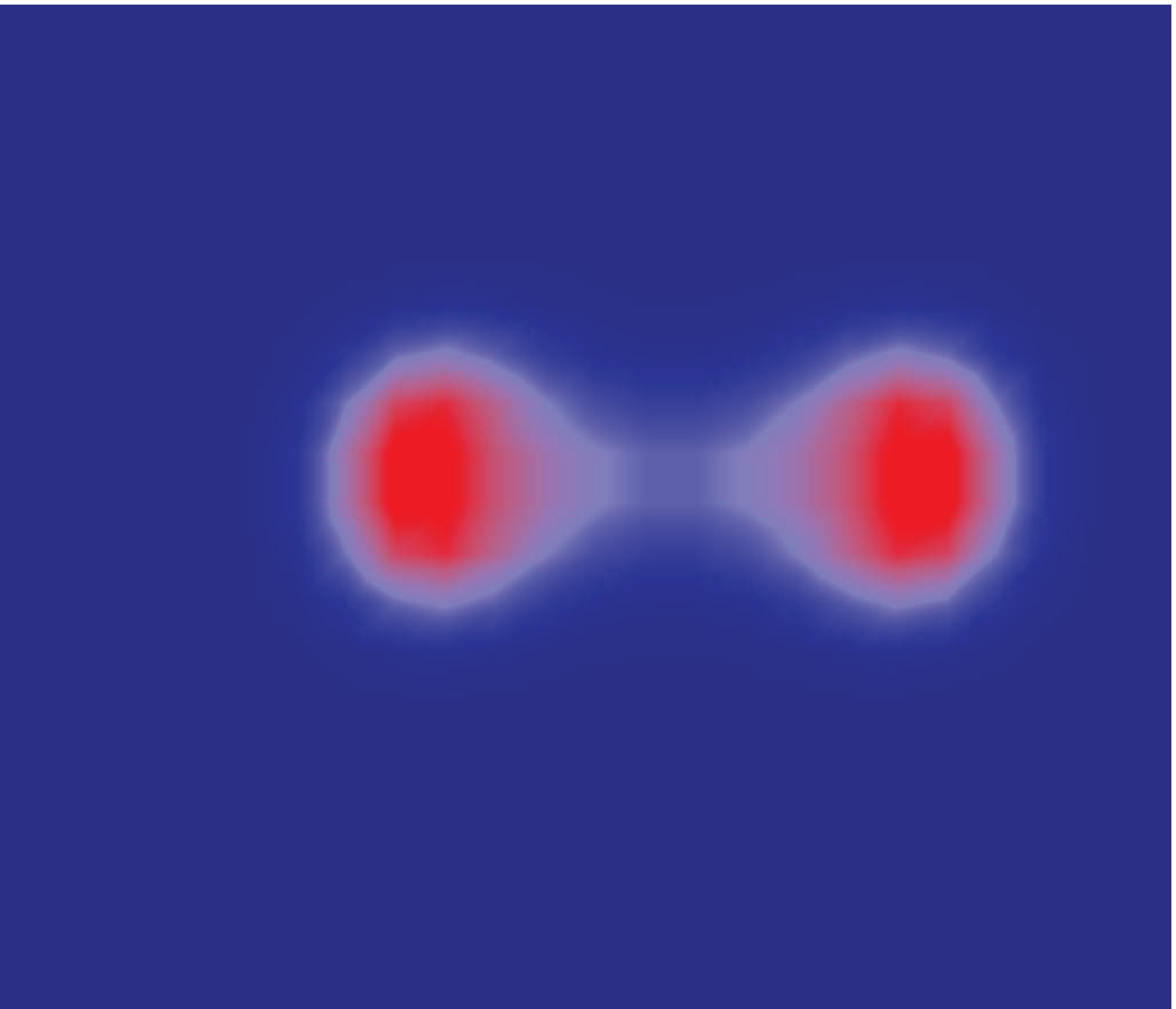}\\[-36pt]
\includegraphics[width=120pt]{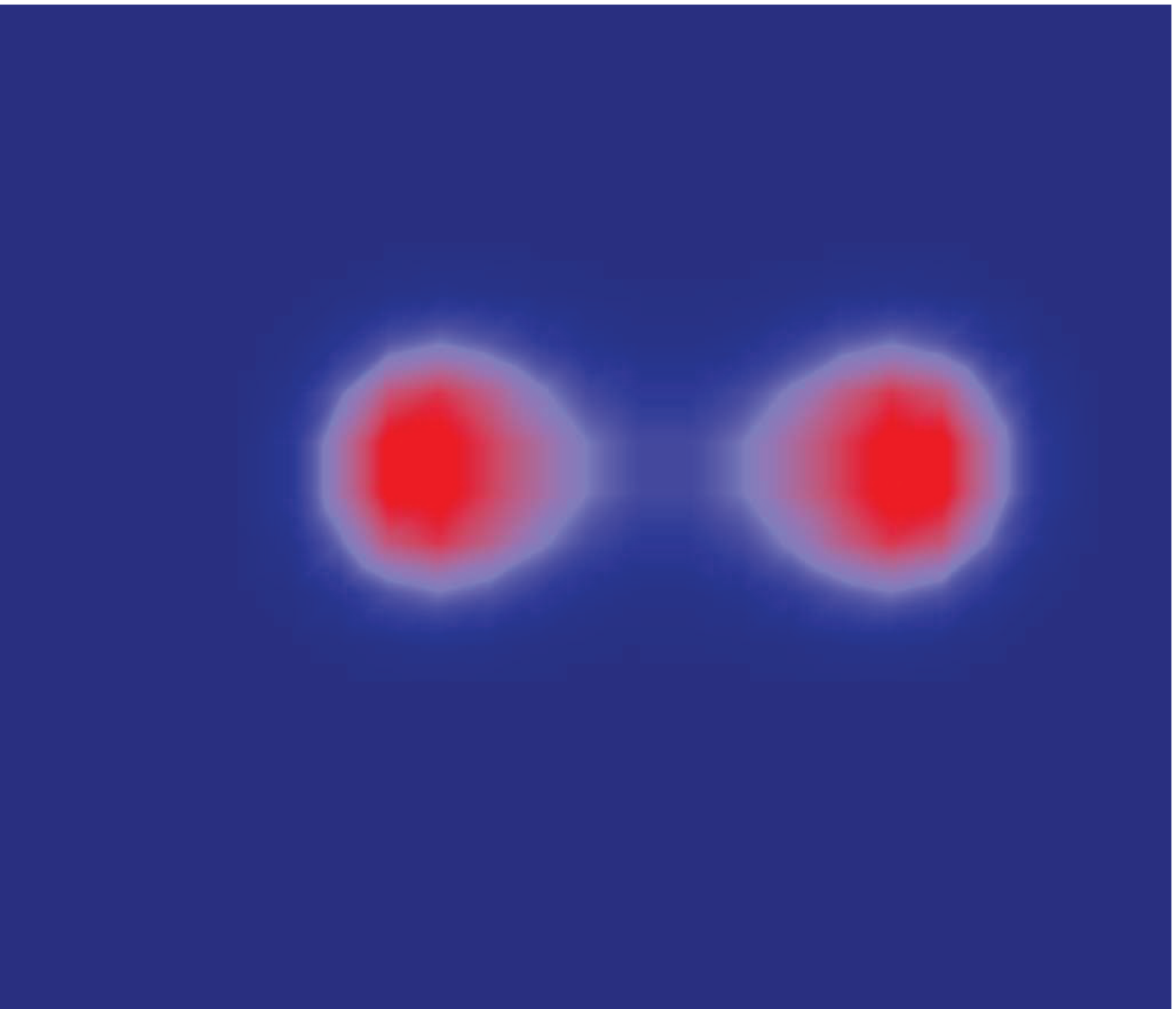}\\
  \caption{(Color online) Density profile of the separating ions at the same relative distance $R=8.3$ fm for three incident
  energies $E_{\rm c.m.}=$ 90 MeV (top), 130 MeV (middle), and 170 MeV (bottom) for the head-on collisions of $^{16}$O+$^{16}$O with the full l*s force.}
  \label{Dens}
\end{figure}

\label{modification:result mechanism}
In order to understand the systematic variation of energy dissipation as the increase of incident energy, we define a quantity which measures the extent of
 energy dissipation in deep-inelastic collisions as $P_{\rm dis}=1-E_{\rm fin}/E_{\rm c.m.}$, where $E_{\rm fin}$ and
${E_{\rm c.m.}}$ denote, respectively, the final and initial kinetic energy already shown in Fig.~\ref{Efin}. The percentage of energy dissipation as a function of incident energy is plotted in Fig.~\ref{Pdis} for head-on collisions of $^{16}$O+$^{16}$O with parametrization SLy4. The black, red, and blue lines represent the TDHF calculations involving full
l*s, time-even l*s, and no l*s force in Skyrme EDF. As can be seen, the inclusion of l*s force results in a significant enhancement of energy dissipation in deep-inelastic collisions. For all three cases the percentage of energy dissipation decreases as the incident energy increases.
To clarify the dissipation mechanism, Fig.~\ref{Dens} displays the density distribution of separating ions
at the same relative distance of 8.3 fm for three incident energies of 90, 130, and 170 MeV in the upper, middle, and bottom panels. The three energies
locate around the start, middle, and end point of the black line shown in Fig.~\ref{Pdis}.
As the increase of incident energy the nucleon density around the neck region becomes lower and at c.m. energy of 170 MeV the nuclei are essentially spherical without a neck formation, indicating that fewer nucleons have been transferred
between the colliding partners. The dynamical mechanism in the exit channel of heavy-ion scattering visualized in Fig.~\ref{Dens} may be interpreted by the concept of dissipative diabatic dynamics which was introduced in the 1980s~\cite{Norenberg1981_PLB104-107} and later applied in the fusion of heavy nuclei~\cite{DiazTorres2000_PLB481-228}. As the increase of incident energy, the collective
motion is so fast that the mean-field does not have enough time to rearrange itself and has to keep its identity as much as possible.
This energy dependence of density evolution has also been presented in the entrance channel of fusion reactions at lower energies for the same and heavier systems~\cite{Washiyama2008_PRC78-024610,Washiyama2009_PRC79-024609,Umar2014_PRC89-034611}.

\begin{figure}
  \includegraphics[width=280pt]{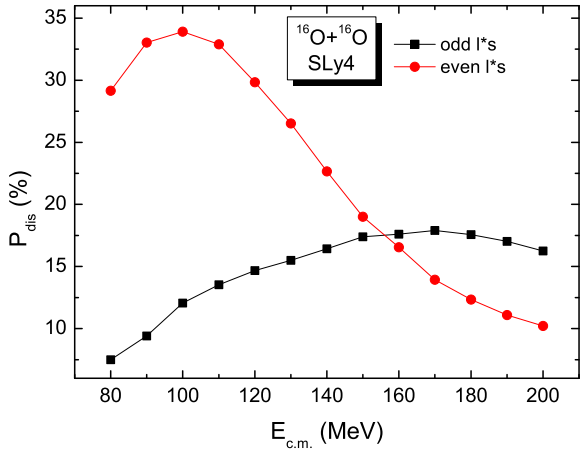}\\
  \caption{(Color online) Percentage of energy dissipation caused by the time-odd and time-even contributions of l*s force as a function of center-of-mass energy for head-on collisions of $^{16}$O+$^{16}$O with parametrization SLy4.}
  \label{Pls}
\end{figure}

\label{modification:result MeVpernucleon}
The percentage of energy dissipation arising solely from the time-even and time-odd contributions of l*s force are shown in Fig.~\ref{Pls}.
The energy dissipation caused by the time-odd l*s force is equal to the percentage of energy dissipation with TDHF
calculations involving full l*s force minus that with time-even l*s, which can be extracted from the black and red lines in Fig.~\ref{Pdis}.
The dissipation from the time-even contribution is the difference between $P_{\rm dis}$ with even l*s and without l*s. As seen in Fig.~\ref{Pls}, the time-even l*s force provides more important
contributions to the dissipation than the time-odd terms when the relative c.m. energy is less than 160 MeV for the $^{16}$O+$^{16}$O collisions.
At low energies the time-even l*s force plays a dominant role and the effect of time-odd l*s becomes more pronounced at high energies.
This might be the reason that earlier TDHF calculations at low collision
energy~\cite{Umar1986_PRL56-2793}, though the time-odd l*s force was neglected, can account for, to some extent, the experimental data~\cite{Lazzarini1981_PRC24-309}.
The dissipation from the time-odd contribution increases as the incident energy, as we expected, because the current density appearing in the
time-odd l*s functional is proportional to the velocity for the case of a nucleus moving with constant velocity.

\begin{figure}
  \includegraphics[width=280pt]{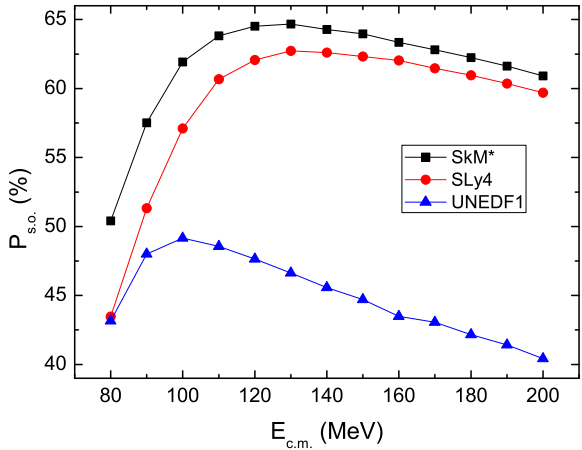}\\
  \caption{(Color online) Percentage of dissipated energy arising from l*s force as a function of center-of-mass energy for head-on collisions of $^{16}$O+$^{16}$O with parametrizations SkM*, SLy4, and UNEDF1.}
  \label{Pso}
\end{figure}

\label{modification:result s.o.}
A quantity which expresses the proportion of dissipated energy coming from l*s force is defined as
$P_{\rm s.o.}=1-P_{\rm dis}^{({\rm no\ l*s})}/P_{\rm dis}^{({\rm full\ l*s})}$, where $P_{\rm dis}^{({\rm no\ l*s})}$ and $P_{\rm dis}^{({\rm full\ l*s})}$ correspond to the blue and black lines in Fig.~\ref{Pdis}. The results are displayed in Fig.~\ref{Pso} for head-on collisions of $^{16}$O+$^{16}$O with the parametrizations SkM*, SLy4, and UNEDF1.
The percentage of dissipated energy coming from l*s force has similar trend for SkM* and SLy4, but evident distinction appears for UNEDF1.
As the increase of incident energy, the dissipation from l*s force increases and then start to decrease after reaching the maximum value of 65\% for SkM*, 63\% for SLy4, and 48\% for UNEDF1. The UNEDF1 calculations predict an overall smaller dissipation arising from l*s force and lower incident energy at the peak of dissipation due to the largest total dissipation (denominator in the definition of $P_{\rm s.o.}$) as seen in Fig.~\ref{Efin} and the smallest dissipated energy from l*s force (numerator in $P_{\rm s.o.}$). The new UNEDF1 parametrization was obtained by additional fitting to the experimental excitation
energies of fission isomers in the actinides and is suitable for studies of strongly deformed nuclei~\cite{Kortelainen2012_PRC85-024304}. The origin of the discrepancy between UNEDF1 results and those with SkM* and SLy4 parameter sets is still unclear.
In spite of the sensitivity of force parameters, it is impressed that more than half of the total dissipation for SkM* and SLy4, and slightly less than half for UNEDF1 are attributed to the l*s force.

\label{modification:result fusion}
We also performed fusion calculations for the $^{16}$O+$^{16}$O system at a c.m. energy of 70.5 MeV. The reason for choosing this collision energy is due to
the availability of experimental data at the energy range studied here. The fusion cross section using three Skyrme
parametrizations at a c.m. energy of 70.5 MeV is listed in Table~\ref{tab:fusion}, together the available experimental cross section with errors~\cite{SaintLaurent1979_327-517}.
TDHF calculations with SkM* and UNEDF1 parametrizations predict the central fusion collisions at the energy of 70.5 MeV because the
high-energy fusion threshold, as shown in Fig.~\ref{Efin}, is higher than 70.5 MeV.
For SLy4 calculations the central transparency predicted by early TDHF calculations~\cite{Davies1980_PRL44-23,Devi1981_PRC23-1064,Maruhn1985_PRC31-1289} appears and the lower limit of angular momentum will become nonzero.
Hence, in the SLy4 calculation of fusion cross section,
we search both the maximum and minimum impact parameter $b_{\rm max}$ and $b_{\rm min}$ at which fusion happens.
Here the fusion is defined as the compound system evolves long enough to offer convincing evidence that the system will not
undergo separation. The fusion window for angular momentum is found to be $b_{\rm min}$=1.2 fm and $b_{\rm max}$=6.5 fm for SLy4,
$b_{\rm max}$=6.5 fm for SkM*, and 6.4 fm for UNEDF1 within a precision of 0.1 fm.
The fusion cross section obtained with the three parametrizations overestimates the experimental value by about 20\%, which is a reasonable agreement considering that no free parameters are adjusted for the reaction dynamics in TDHF calculations.
The cross section for transparency at 70.5 MeV is predicted to be quite small with 45 mb for SLy4. Whether this effect is
real or not would not change much the conclusion that the fusion cross section is reasonably reproduced by TDHF calculations at this energy.

\label{modification:result figure}

\label{modification:result fusiondynamics}
The heavy-ion fusion at energies of several times the barrier height is a challenge both for theoretical and experimental studies. As shown by Simenel {\it et al.}~\cite{Simenel2013_PRC88-024617}, for the $^{16}$O+$^{16}$O system at $E_{\rm c.m.}$=35 MeV, the compound nucleus cannot achieve a complete thermal equilibration between collective motion and internal excitation after the long-time evolution and finally settles at an energy difference of about 10 MeV between the c.m. energy and the excitation energy. Due to the large remaining collective kinetic energy, the compound system is likely to undergo various breakup and fission events which are beyond the description of TDHF theory. Since these events are regarded as fusion in TDHF, a larger fusion cross section is expected at high energies in
TDHF calculations. In addition, two-body collisions which are likely to play an important role at high energies have been neglected in mean-field dynamics. From the experimental point of view,
the fusion cross sections for the $^{16}$O+$^{16}$O system have been measured by several groups using different experimental approaches and techniques~\cite{Tserruya1978_PRC18-1688, Fernandez1978_NPA306-259, Kolata1979_PRC19-2237}. The different sets of
experimental data are not consistent with each other at energies of two to four times barrier height, leading to a large uncertainty on the experimental cross sections~\cite{Simenel2013_PRC88-024617}.
The experimental cross section at c.m. energy of 70.5 MeV, although only one set data is available, is expected to have large variations as the cases at lower energies. Reliable experimental data at well above the barrier energy are highly desired.

The dissipation dynamics has also been investigated for medium-mass system of $^{40}$Ca+$^{40}$Ca in order to see the systematic variations. There is no qualitative
distinction compared with the light system of $^{16}$O+$^{16}$O. Therefore, the results for heavier system will not be incorporated in this article.

\label{modification:result table}
\begin{table}
\caption{Calculations of fusion cross section for $^{16}$O+$^{16}$O at $E_{\rm{c.m.}}=70.5$ MeV with three Skyrme parametrizations and
experimental data with errors~\cite{SaintLaurent1979_327-517}.}
\label{tab:fusion}
\begin{center}
\begin{tabular*}{60mm}{@{\extracolsep{\fill}}lr}
\hline
\hline
  Force    & $\sigma_{\rm{fus}}$ (mb)      \\
\hline
SLy4  &    1282         \\
SkM*  &    1287         \\
UNEDF1 &   1327        \\
Expt.  &    $1056\pm125$    \\
\hline
\hline
\end{tabular*}
\end{center}
\end{table}

\section{Summary}
\label{summary}
We have investigated the dissipation dynamics in heavy-ion collisions of $^{16}$O+$^{16}$O using the modern TDHF approach. The numerical
calculations have been implemented in a fully three-dimensional coordinate space with the modern Skyrme energy functional and without any symmetry restrictions.
The dissipation dynamics exhibits a universal behavior by using three Skyrme parametrizations SkM*, SLy4, and UNEDF1.
We revealed that
the percentage of energy dissipation in deep-inelastic collisions of the light systems decreases as the increase of incident energy due to the interplay between the
collective motion and the single particle degrees of freedom.
TDHF calculations with the newly fitted parametrization UNEDF1 predict the largest energy dissipation, while the smallest with SLy4 among the three parametrizations. The energy
dissipation in the present studies has been compared with the results in earlier calculations.  Special attention is paid to the role of time-even and time-odd spin-orbit force.
The spin-orbit force significantly enhances the energy dissipation.
The time-even coupling of spin-orbit force plays a dominant role at low energies, while the effect of time-odd terms becomes more pronounced at high
energies. More than half of the total dissipation for SkM* and SLy4, and slightly less than half for UNEDF1 are predicted to come from spin-orbit force. We also performed the fusion calculations for this system. The theoretical fusion cross section obtained from the parameter-free TDHF approach agrees reasonably well with the experimental data.

\section{Acknowledgments}
This work was supported by Natural Science Foundation of China (Grants No. 11175252, No. 11121403, No. 11120101005, No. 11211120152, and No. 11275248),
the National Key Basic Research Program of China (Grant No. 2013CB834400), the Knowledge Innovation Project of the Chinese Academy of Sciences (Grant No. KJCX2-EW-N01), the President Fund of UCAS,
the Scientific Research Foundation for the Returned Overseas Chinese Scholars, Ministry of Education of China, and
the Open Project Program of State Key Laboratory of Theoretical Physics,
Institute of Theoretical Physics, Chinese Academy of Sciences, China (Grant No.Y4KF041CJ1).
The results described in this paper were obtained on the High-performance Computing Clusters of SKLTP/ITP-CAS and the ScGrid of the Supercomputing Center, Computer Network Information Center of the Chinese Academy of Sciences.

\bibliographystyle{apsrev4-1}
\bibliography{ref}

\end{document}